\documentclass[11pt,executivepaper]{article}
\usepackage{amssymb}
\usepackage{amsmath}

\input{tcilatex}

\begin{document}

\begin{center}
{\Large Generalized Morse and P\"{o}schl-Teller potentials :}

{\Large The connection via Schr\"{o}dinger equation}

\bigskip 

S.-A. Yahiaoui, S. Hattou, M. Bentaiba.\footnote{%
Corresponding author:
\par
E-mail address : bentaiba@hotmail.com}

LPTHIRM, D\'{e}partement de Physique, Facult\'{e} des Sciences,

Universit\'{e} Saad DAHLAB de Blida, Alg\'{e}rie.

\bigskip 

\bigskip 

\textbf{Abstract}
\end{center}

We present here a systematic and unified treatment to connect the
Schrodinger equation corresponding to generalized Morse and Poschl-Teller
potentials. We then show that the wave functions and generalized potentials
are linked through the Fourier and Hankel transforms, respectively.

\bigskip 

PACS: 03.65.Ca; 03.65.-w; 03.65.Fd; 02.30.Nw.

KeyWords: Formalism, Algebraic methods, Group theory, Fourier analysis.

\bigskip 

\section{Introduction}

Group theory provides us with efficient algebraic techniques which are used
for the description of the energy spectra of quantum mechanical potentials
[1-5]. One of the important aspects of algebraic techniques is how to
construct the Hamiltonian from the Casimir operator(s) related to the
group-algebraic structures. To this end, the most used technique is the
potential group approach [1,4,5].\ It's based on two approaches, the former,
called the algebraic approach [1], is considered as the convenient way to
construct the spectrum-generating algebra for the system and this by
introducing a set of boson creation and annihilation operators. Therefore,
it can be connected to the exactly soluble Schr\"{o}dinger equations with
certain potentials, whereas the latter called potential approach [1] can be
connected to the first one, i.e. algebraic approach, by introducing the
boson creation and annihilation operators as differential operators in
two-dimensional harmonic oscillator space and sphere.

The supersymmetric quantum mechanics is an another algebraic approach based
on the generalized version of the creation and annihilation operators
provided by the factorization method and investigates Hamiltonians which
split into two components whose spectra are the same, with the exception of
the ground-state [6-14]. These operators are often called the bosonic and
fermionic parts of supersymmetric system.

\ We concentrate our attention in this paper on the Morse and the P\"{o}%
schl-Teller one-dimensional potentials. The former can be realized on $%
SU\left( 2\right) $\ two-dimensional harmonic oscillator space [15], whereas
the latter can be realized on $SU\left( 2\right) $-sphere [16]. It has been
proved that the energy spectra of both potentials are related to the same
representation of the $\mathfrak{su}\left( 2\right) \backsimeq \mathfrak{so}%
\left( 3\right) $ Lie algebra ($\backsimeq $\ means isomorphic to) [1]; it
plays the role of dynamical algebra. Therefore, the $\mathfrak{su}\left(
2\right) $\ Lie algebra can gives unitary representation of the $SU\left(
2\right) $\ Lie group.

The connection between the Schr\"{o}dinger equations with Morse and P\"{o}%
schl-Teller potentials was already established by Alhassid et al. [1]. This
present paper would be an extension of algebraic treatment\ of $\mathfrak{su}%
\left( 2\right) $ Lie Algebra to their generalized potentials, through the
supersymmetric quantum mechanics which allows us to generalize any given
potential, with a view to establishing a causal connection underlying the
relationship between both generalized Morse and P\"{o}schl-Teller
potentials. We then show that the Schr\"{o}dinger equation for generalized P%
\"{o}schl-Teller potentials can be constructed mathematically starting off
with a Schr\"{o}dinger equation for generalized Morse potentials, proving
that each equation can be interpreted as the Fourier transform of the other
(see the appendix). As a consequence of this, the algebra associated with $%
SU\left( 2\right) $ group provides, in a straightforward way, the relation
connecting both energy spectra.

The arrangement of this paper is as follows. First we recall a brief
overview of $\mathfrak{su}\left( 2\right) $ Lie algebra and the basic
relations involved in supersymmetric quantum mechanics, in order to
construct the generalized Morse [13] and P\"{o}schl-Teller potentials.
Section 3 deals with different steps of transformation connecting\ their
generalized potentials via the Schr\"{o}dinger equation. The final section
will be devoted to discussions and in the appendix, mathematical details of
transformation connecting generalized potentials will be presented.

\section{Classes of $SU\left( 2\right) $ realizations and Supersymmetry}

The generators of the compact $SU\left( 2\right) $ Lie group obey to
commutation relations [1]%
\begin{eqnarray}
\left[ \mathcal{J}_{z},\mathcal{J}_{\pm }\right] &=&\pm \mathcal{J}_{\pm }, 
\TCItag{1.a} \\
\left[ \mathcal{J}_{+},\mathcal{J}_{-}\right] &=&2\mathcal{J}_{z}. 
\TCItag{1.b}
\end{eqnarray}%
with $\mathcal{J}_{\pm }=\mathcal{J}_{x}\pm i\mathcal{J}_{y}$. Here, the
operators$\ \mathcal{J}_{\alpha }$ with $\alpha =+,-,z$\ can be obtained by
using the set of boson operators as%
\begin{eqnarray}
\mathcal{J}_{+} &=&a^{\dag }b,  \TCItag{2.a} \\
\mathcal{J}_{-} &=&b^{\dag }a,  \TCItag{2.b} \\
\mathcal{J}_{z} &=&\frac{1}{2}\left( a^{\dag }a-b^{\dag }b\right) . 
\TCItag{2.c}
\end{eqnarray}

In order to complete the $SU\left( 2\right) $\ algebra, one needs a fourth
operator namely the total boson number operator\ $\mathcal{N}=a^{\dag
}a+b^{\dag }b$ which belongs to $U\left( 2\right) $ group, and is related to
the Casimir operator $\mathcal{C}_{2}$\ of $U\left( 2\right) $ by%
\begin{eqnarray}
\mathcal{C}_{2} &=&\frac{1}{4}\mathcal{N}\left( \mathcal{N}+2\right) 
\nonumber \\
&=&\mathcal{J}_{+}\mathcal{J}_{-}+\mathcal{J}_{z}\left( \mathcal{J}%
_{z}-1\right) .  \TCItag{3}
\end{eqnarray}

Therefore, it is clear that the eigenstates of $\mathcal{C}_{2}$\ and $%
\mathcal{J}_{z}$\ serve as basis for the irreducible representation of $%
\mathfrak{su}\left( 2\right) $\ algebra. Let us derive now the standard
supersymmetric quantum mechanics (SUSY-QM) relations. Considering SUSY-QM
with $\mathfrak{D}=2$ [6-14], we define the supercharges $\mathcal{Q}%
=d\sigma _{-}$ and $\mathcal{Q}^{\dag }=d^{\dag }\sigma _{+}$\ where $%
d\left( d^{\dag }\right) $\ and $\sigma _{-}\left( \sigma _{+}\right) $\
are, respectively, the bosonic operators and Pauli matrices%
\begin{eqnarray}
d &=&\partial _{x}+x,  \TCItag{4.a} \\
d^{\dag } &=&-\partial _{x}+x.  \TCItag{4.b}
\end{eqnarray}

\begin{equation}
\sigma _{-}=%
\begin{pmatrix}
0 & 0 \\ 
1 & 0%
\end{pmatrix}%
\quad ;\quad \sigma _{+}=%
\begin{pmatrix}
0 & 1 \\ 
0 & 0%
\end{pmatrix}%
.  \tag{5}
\end{equation}%
where we have used the abbreviation $\partial _{x}=\dfrac{d}{dx}$.

\ We define the supersymmetric Hamiltonian [11,13]%
\begin{equation}
\mathcal{H}_{SUSY}\equiv \left\{ \mathcal{Q},\mathcal{Q}^{\dag }\right\} 
\footnote[1]{%
Here $\left\{ \cdots ,\cdots \right\} $ stands for the anticommutation
relations, i.e. $\left\{ A,B\right\} =AB+BA.$ }=%
\begin{pmatrix}
\mathcal{H}_{+} & 0 \\ 
0 & \mathcal{H}_{-}%
\end{pmatrix}%
=%
\begin{pmatrix}
d^{\dag }d & 0 \\ 
0 & dd^{\dag }%
\end{pmatrix}%
,  \tag{6}
\end{equation}%
where $\mathcal{H}_{-}$\ is called the supersymmetric partner of $\mathcal{H}%
_{+}$. Both $\mathcal{H}_{+}$ and $\mathcal{H}_{-}$\ have the same spectra
except for the ground-state, which belongs to $\mathcal{H}_{+}$.

In order to find the generalized potentials, we shall first generalize the
bosonic operators $d$\ and $d^{\dag }$\ [6,11] as%
\begin{eqnarray}
\mathcal{D} &=&\partial _{x}+f\left( x\right) ,  \TCItag{7.a} \\
\mathcal{D}^{\dag } &=&-\ \partial _{x}+f\left( x\right) .  \TCItag{7.b}
\end{eqnarray}

By imposing that $\mathcal{H}_{-}=dd^{\dag }$, we obtain the Ricatti
differential equation from the Schr\"{o}dinger equation%
\begin{equation}
\partial _{x}f\left( x\right) +f^{2}\left( x\right) =\left( \partial
_{x}W\left( x\right) \right) ^{2}+\partial _{x}^{2}W\left( x\right) , 
\tag{8}
\end{equation}%
where $f\left( x\right) $ is the derivative of $W\left( x\right) $, this
later\ is called superpotential and is associated with $\mathcal{H}_{+}$\
ground-state eigenfunctions $\psi _{+,0}$.

Therefore, we can define a new Hamiltonian using (7)%
\begin{equation}
\widetilde{\mathcal{H}}_{+}=\mathcal{D}^{\dag }\mathcal{D}=\mathcal{DD}%
^{\dag }-\left[ \mathcal{D},\mathcal{D}^{\dag }\right] =\mathcal{DD}^{\dag
}-2\ \partial _{x}W\left( x\right) .  \tag{9}
\end{equation}

Following supersymmetry, it seems that the spectrum of $\widetilde{\mathcal{H%
}}_{+}$\ is the same as the spectrum of $\mathcal{H}_{-}$ [8,11].

\subsection{The first class $SU\left( 2\right) $ realization : Morse
potential}

The algebraic approach based on (1)-(3) can be connected to the Schr\"{o}%
dinger equation of the Morse potential into a two-dimensional harmonic space
by introducing the following realizations of the boson operators [1]%
\begin{eqnarray}
a &=&\dfrac{1}{\sqrt{2}}\left( x+\partial _{x}\right) ,  \TCItag{10.a} \\
a^{\dag } &=&\dfrac{1}{\sqrt{2}}\left( x-\partial _{x}\right) , 
\TCItag{10.b} \\
b &=&\dfrac{1}{\sqrt{2}}\left( y+\partial _{y}\right) ,  \TCItag{10.c} \\
b^{\dag } &=&\dfrac{1}{\sqrt{2}}\left( y-\partial _{y}\right) . 
\TCItag{10.d}
\end{eqnarray}

In terms of the variables $x,\ y$, the two operators $\mathcal{N}$\ and $%
\mathcal{J}_{z}$\ become%
\begin{eqnarray}
\mathcal{N} &=&\dfrac{1}{2}\left( x^{2}+y^{2}-\partial _{x}^{2}-\partial
_{y}^{2}-2\right) ,  \TCItag{11.a} \\
\mathcal{J}_{z} &=&-\dfrac{i}{2}\left( x\ \partial _{x}-y\ \partial
_{y}\right) .  \TCItag{11.b}
\end{eqnarray}

A change of variables $x=r\cos \varphi ,\ y=r\sin \varphi $,\ with $0\leq
r<\infty $\ and $0\leq \varphi <2\pi $ would be helpful,\ leading to
re-express (11.a) and (11.b) by%
\begin{eqnarray}
\mathcal{N} &=&\dfrac{1}{2}\left( -\frac{1}{r}\partial _{r}\left( r\partial
_{r}\right) -\frac{1}{r^{2}}\partial _{\varphi }^{2}+r^{2}\right) -1, 
\TCItag{12.a} \\
\mathcal{J}_{z} &=&-\dfrac{i}{2}\ \partial _{\varphi }.  \TCItag{12.b}
\end{eqnarray}

The action of (12.a) on the wave function $\psi \left( r,\varphi \right) =%
\mathcal{R}\left( r\right) \func{e}^{2\sqrt{E_{M}}\varphi }$, taking into
account the transformation $r^{2}=2\lambda \func{e}^{-\rho }$, leads to the
Schr\"{o}dinger equation for a particle constrained to move in
one-dimensional Morse potential%
\begin{equation}
\left[ -\partial _{\rho }^{2}+\lambda ^{2}\left( \func{e}^{-2\rho }-2\func{e}%
^{-\rho }\right) \right] \mathcal{R}\left( r\right) =E_{M}\mathcal{R}\left(
r\right) .  \tag{13}
\end{equation}

In order to make the eigenvalues of the ground-state equal to zero, i.e. $%
E_{+,0}^{\left( M\right) }=0$, one displaces the potential given in (13) by
the quantity $\left( \lambda -\frac{1}{2}\right) ^{2}$, i.e.

\begin{eqnarray}
\lambda ^{2}\left( \func{e}^{-2\rho }-2\func{e}^{-\rho }\right)
&\longrightarrow &\lambda ^{2}\left( \func{e}^{-2\rho }-2\func{e}^{-\rho
}\right) +\left( \lambda -\frac{1}{2}\right) ^{2}  \nonumber \\
&=&\lambda ^{2}\left( 1-\func{e}^{-\rho }\right) ^{2}-\lambda +\frac{1}{4}. 
\TCItag{14}
\end{eqnarray}

As a consequence of this, the Hamiltonian $\mathcal{H}_{+}$ given in (6)
becomes [13]%
\begin{equation}
\mathcal{H}_{+}\equiv d^{\dag }d=-\ \partial _{\rho }^{2}+\lambda ^{2}\left(
1-\func{e}^{-\rho }\right) ^{2}-\lambda +\frac{1}{4},  \tag{15}
\end{equation}%
and then the supersymmetric partner of the Hamiltonian (15) reads

\begin{equation}
\mathcal{H}_{-}\equiv d\ d^{\dag }=\mathcal{H}_{+}+2\ \lambda \func{e}%
^{-\rho }.  \tag{16}
\end{equation}

Both Hamiltonians have the same spectra, except that (15) has a ground-state
with zero eigenvalue.

In order to generalize (15), we define the operators given by (7) and by
solving the corresponding Ricatti differential equation, we obtain%
\begin{equation}
f_{M}\left( \rho \right) =\lambda \left( 1-\func{e}^{-\rho }\right) -\frac{1%
}{2}+\frac{\exp \left[ -\left( 2\lambda -1\right) \rho -2\lambda \func{e}%
^{-\rho }\right] }{\Gamma +\int\limits_{0}^{\rho }d\widetilde{\rho }\exp %
\left[ -\left( 2\lambda -1\right) \widetilde{\rho }-2\lambda \func{e}^{-%
\widetilde{\rho }}\right] }.  \tag{17}
\end{equation}%
where $\Gamma $\ is an arbitrary constant and\ is chosen $\Gamma >0$ in
order to avoid singularities.

We can define a new Hamiltonian $\widetilde{\mathcal{H}}_{+}$\ where the
corresponding potential is%
\begin{equation}
\widetilde{\mathcal{H}}_{+}\left( \rho \right) =-\partial _{\rho
}^{2}+\lambda ^{2}\left( 1-\func{e}^{-\rho }\right) ^{2}-\lambda +\frac{1}{4}%
-2\partial _{\rho }\left[ \frac{\exp \left[ -\left( 2\lambda -1\right) \rho
-2\lambda \func{e}^{-\rho }\right] }{\Gamma +\int\limits_{0}^{\rho }d%
\widetilde{\rho }\exp \left[ -\left( 2\lambda -1\right) \widetilde{\rho }%
-2\lambda \func{e}^{-\widetilde{\rho }}\right] }\right] .  \tag{18}
\end{equation}

The equation (18) is different from the original potential (15) and also
from $\mathcal{H}_{-}$\ given in (16), but we know from supersymmetry that
the spectrum of $\widetilde{\mathcal{H}}_{+}$\ exhibits the same energy
spectrum as $\mathcal{H}_{-}$\ except for the ground-state [13].

The Schr\"{o}dinger equation corresponding to generalized Morse potential
(18) is%
\begin{equation}
\left[ -\partial _{\rho }^{2}+\lambda ^{2}\left( 1-\func{e}^{-\rho }\right)
^{2}-\lambda +\frac{1}{4}-2\partial _{\rho }q_{\lambda ,\Gamma }^{\left(
M\right) }\left( \rho \right) \right] \mathcal{R}\left( \rho \right) =E_{M}%
\mathcal{R}\left( \rho \right) ,  \tag{19}
\end{equation}%
with%
\begin{equation}
q_{\lambda ,\Gamma }^{\left( M\right) }\left( \rho \right) =\frac{\exp \left[
-\rho \left( 2\lambda -1\right) -2\lambda \func{e}^{-\rho }\right] }{\Gamma
+\int\limits_{0}^{\rho }d\widetilde{\rho }\exp \left[ -\widetilde{\rho }%
\left( 2\lambda -1\right) -2\lambda \func{e}^{-\widetilde{\rho }}\right] }. 
\tag{20}
\end{equation}

The change of variable $\rho =\ln \frac{2\lambda }{r^{2}}$ and the
introduction of a parameter $a=\lambda -\frac{1}{2}$ bring (19) and (20) to%
\begin{equation}
\left[ -\frac{1}{r}\partial _{r}\left( r\partial _{r}\right) +r^{2}+\frac{%
4\left( a^{2}-E_{M}\right) }{r^{2}}+\frac{4}{r}\partial _{r}q_{a,\Gamma
}^{\left( M\right) }\left( r\right) \right] \mathcal{R}\left( r\right)
=\left( 4a+2\right) \mathcal{R}\left( r\right) ,  \tag{21}
\end{equation}%
with%
\begin{equation}
q_{a,\Gamma }^{\left( M\right) }\left( r\right) =\frac{\left( \dfrac{r^{2}}{%
2a+1}\right) ^{2a}\func{e}^{-r^{2}}}{\Gamma +\int\limits_{0}^{-\ln
r^{2}/2a+1}d\widetilde{r}\left( \dfrac{-2}{\widetilde{r}}\right) \left( 
\dfrac{\widetilde{r}^{2}}{2a+1}\right) ^{2a}\func{e}^{-\widetilde{r}^{2}}} 
\tag{22}
\end{equation}

\subsection{The second class $SU\left( 2\right) $ realization : P\"{o}%
schl-Teller potential}

A realization of $\mathfrak{su}\left( 2\right) $\ algebra on the sphere
leads to a connection between $\mathfrak{su}\left( 2\right) $\ algebra and
the Schr\"{o}dinger equation with P\"{o}schl-Teller potential. This last is
obtained by using the following realizations [1]%
\begin{eqnarray}
\mathcal{J}_{\pm } &=&\func{e}^{\pm i\varphi }\left( \pm \partial _{\theta
}+i\cot \varphi \partial _{\varphi }\right) ,  \TCItag{23.a} \\
\mathcal{J}_{z} &=&-i\partial _{\varphi },  \TCItag{23.b} \\
\mathcal{N} &=&-\left[ \dfrac{1}{\sin \theta }\partial _{\theta }\left( \sin
\theta \partial _{\theta }\right) +\dfrac{1}{\sin ^{2}\theta }\partial
_{\varphi }^{2}\right] .  \TCItag{23.c}
\end{eqnarray}

The substitution $\cos \theta =\tanh \rho $\ with $-\infty <\rho <\infty $\
brings the eigenstate equation $\mathcal{N\xi }\left( \theta ,\varphi
\right) =\mu \left( \mu +1\right) \xi \left( \theta ,\varphi \right) $, with 
$\xi \left( \theta ,\varphi \right) =\mathcal{U}\left( \theta \right) \func{e%
}^{\sqrt{E_{PT}}\varphi }$,\ to a dimensionless Schr\"{o}dinger equation
with P\"{o}schl-Teller potential [1,16]%
\begin{equation}
\left[ -\partial _{\rho }^{2}-\frac{\mu \left( \mu +1\right) }{\cosh
^{2}\rho }\right] \mathcal{U}\left( \rho \right) =E_{PT}\mathcal{U}\left(
\rho \right) .  \tag{24}
\end{equation}

In order to displace the ground-state energy to zero, we add a constant term 
$\mu ^{2}$\ to the potential in (24) such that the Hamiltonian becomes%
\begin{equation}
\mathcal{H}_{+}=-\partial _{\rho }^{2}-\frac{\mu \left( \mu +1\right) }{%
\cosh ^{2}\rho }+\mu ^{2}.  \tag{25}
\end{equation}

The supersymmetric partner of $\mathcal{H}_{+}$, following (6), reads as%
\begin{equation}
\mathcal{H}_{-}=\mathcal{H}_{+}+\frac{2\mu }{\cosh ^{2}\rho }.  \tag{26}
\end{equation}

As in case of Morse potential, $\mathcal{H}_{+}$\ and $\mathcal{H}_{-}$\
share the same spectrum except that (25) has ground-state energy equal to
zero [1-7,17,18].

Performing similar transformations to (7), we define the corresponding
Ricatti differential equation having the solution%
\begin{equation}
f_{PT}\left( \rho \right) =\mu \tanh \rho +\frac{\cosh ^{-2\mu }\rho }{%
\Gamma +\int\limits_{0}^{\rho }d\widetilde{\rho }\cosh ^{-2\mu }\widetilde{%
\rho }},  \tag{27}
\end{equation}%
where $\Gamma >0$\ is added in order to avoid contingent singularities.

Therefore, we can define a new Hamiltonian%
\begin{equation}
\widetilde{\mathcal{H}}_{+}=-\partial _{\rho }^{2}-\frac{\mu \left( \mu
+1\right) }{\cosh ^{2}\rho }+\mu ^{2}-2\partial _{\rho }\left[ \frac{\cosh
^{-2\mu }\rho }{\Gamma +\int\limits_{0}^{\rho }d\widetilde{\rho }\cosh
^{-2\mu }\widetilde{\rho }}\right] .  \tag{28}
\end{equation}

We can verify that $\widetilde{\mathcal{H}}_{+}$\ and $\mathcal{H}_{-}$\
have the same spectra except, once again, for the ground-state energy.

From (28), the Schr\"{o}dinger equation for generalized P\"{o}schl-Teller
potential reads%
\begin{equation}
\left[ -\partial _{\rho }^{2}-\frac{\mu \left( \mu +1\right) }{\cosh
^{2}\rho }+\mu ^{2}-2\partial _{\rho }\left( \frac{\cosh ^{-2\mu }\rho }{%
\Gamma +\int\limits_{0}^{\rho }d\widetilde{\rho }\cosh ^{-2\mu }\widetilde{%
\rho }}\right) \right] \mathcal{U}\left( \rho \right) =E_{PT}\mathcal{U}%
\left( \rho \right) .  \tag{29}
\end{equation}

The following substitutions%
\begin{eqnarray}
\zeta &=&\sinh \rho ,  \TCItag{30.a} \\
\sigma &=&\cot ^{-1}\zeta ,  \TCItag{30.b} \\
t &=&\tan \frac{\sigma }{2},  \TCItag{30.c}
\end{eqnarray}%
bring (29) to the form

\begin{equation}
\left[ -t\partial _{t}\left( t\partial _{t}\right) -4\mu \left( \mu
+1\right) \frac{t^{2}}{\left( 1+t^{2}\right) ^{2}}+\mu ^{2}+2t\partial
_{t}q_{\mu ,\Gamma }^{\left( PT\right) }\left( t\right) \right] \mathcal{U}%
\left( t\right) =E_{PT}\mathcal{U}\left( t\right) ,  \tag{31}
\end{equation}%
with%
\begin{equation}
q_{\mu ,\Gamma }^{\left( PT\right) }\left( t\right) =\frac{\left( \dfrac{2t}{%
1+t^{2}}\right) ^{2\mu }}{\Gamma +\int\limits_{0}^{-\ln t}d\widetilde{t}%
\left( \dfrac{-2}{1+\widetilde{t}^{2}}\right) \left( \dfrac{2t}{1+\widetilde{%
t}^{2}}\right) ^{2\mu -1}}.  \tag{32}
\end{equation}

Dividing now the whole of (31) by $t^{2}$, and by transposing the energy
term to the left-hand side and the second term of the left member to the
right-hand side, we get the final Schr\"{o}dinger equation%
\begin{equation}
\left[ -\frac{1}{t}\partial _{t}\left( t\partial _{t}\right) -\frac{%
E_{PT}-\mu ^{2}}{t^{2}}+\frac{2}{t}\partial _{t}q_{\mu ,\Gamma }^{\left(
PT\right) }\left( t\right) \right] \mathcal{U}\left( t\right) =\frac{4\mu
\left( \mu +1\right) }{\left( 1+t^{2}\right) ^{2}}\mathcal{U}\left( t\right)
.  \tag{33}
\end{equation}

Since both the Morse and the P\"{o}schl-Teller potentials appear to be
related to the same representations of $SU\left( 2\right) $, we conjecture
that, as was made for potentials [1], there must be a transformation
connecting their generalized potentials (18) and (28), and then their
corresponding Schr\"{o}dinger equation (21) and (33). It is what we are
going to prove in section 3.

\section{Connection between generalized Morse and P\"{o}schl-Teller
potentials\ }

Before proceeding, let us recall that the solutions of the Schr\"{o}dinger
equations with the generalized Morse potentials, namely $\psi \left(
r,\varphi \right) $, should be periodic in $\varphi $\ with period $2\pi $,
and it can be written in the form%
\begin{equation}
\psi \left( r,\varphi \right) =\func{e}^{2im\varphi }\mathcal{R}\left(
r\right) ,  \tag{34}
\end{equation}%
where $m$\ is an integer carrying information about energy.

By performing the second derivative of $\psi \left( r,\varphi \right) $ with
respect to $\varphi $, and identifying the result with the third term in
(21), we get%
\begin{equation}
\partial _{\varphi }^{2}\equiv -4m^{2}=-4\left( a^{2}-E_{M}\right) , 
\tag{35}
\end{equation}%
and which permits to re-express (21) in the form%
\begin{equation}
\left[ -\frac{1}{r}\partial _{r}\left( r\partial _{r}\right) +r^{2}-\frac{1}{%
r^{2}}\partial _{\varphi }^{2}+\frac{4}{r}\partial _{r}q_{a,\Gamma }^{\left(
M\right) }\left( r\right) \right] \psi \left( r,\varphi \right) =\left(
4a+2\right) \psi \left( r,\varphi \right) .  \tag{36}
\end{equation}

Let us make a change of variables by introducing the bidimensional vector $%
\mathbf{t}\equiv \left( t_{x},t_{y}\right) $\ through [1]%
\begin{eqnarray}
t_{x} &=&t\cos \Phi \quad ;\quad t_{y}=t\sin \Phi .  \TCItag{37.a} \\
t &=&\dfrac{r^{2}}{2}\quad \quad \ \ ;\quad \Phi =2\varphi .  \TCItag{37.b}
\end{eqnarray}

By performing the first and second derivatives with respect to $r$\ and $%
\varphi $, taking into account (37), the Schr\"{o}dinger equation (36)
becomes%
\begin{equation}
t\left[ -\frac{1}{t}\partial _{t}\left( t\partial _{t}\right) -\frac{1}{t^{2}%
}\partial _{\Phi }^{2}+\frac{2}{t}\partial _{t}q_{a,\Gamma }^{\left(
M\right) }\left( t\right) +1\right] \psi \left( t,\Phi \right) =\left(
2a+1\right) \psi \left( t,\Phi \right) .  \tag{38}
\end{equation}

The canonically conjugate momenta vector, $\mathbf{\tau }\equiv \left( \tau
_{x},\tau _{y}\right) $\ associated to $\mathbf{t}\equiv \left(
t_{x},t_{y}\right) $, is introduced [1]. Consequently, (38) can be written as%
\begin{equation}
t\left( 1+\tau ^{2}\right) \psi \left( t,\Phi \right) =\left( 2a+1\right)
\psi \left( t,\Phi \right) ,  \tag{39}
\end{equation}%
with%
\begin{equation}
\tau ^{2}\equiv -\frac{1}{t}\partial _{t}\left( t\partial _{t}\right) -\frac{%
1}{t^{2}}\partial _{\Phi }^{2}+\frac{2}{t}\partial _{t}q_{a,\Gamma }^{\left(
M\right) }\left( t\right) .  \tag{40}
\end{equation}

Proceeding now to square (39). This amounts to multiply this later on the
left-hand side by $t\left( 1+\tau ^{2}\right) $, i.e.%
\begin{eqnarray}
t\left( 1+\tau ^{2}\right) t\left( 1+\tau ^{2}\right) \psi \left( t,\Phi
\right) &=&\left( 2a+1\right) t\left( 1+\tau ^{2}\right) \psi \left( t,\Phi
\right)  \nonumber \\
&=&\left( 2a+1\right) ^{2}\psi \left( t,\Phi \right) .  \TCItag{41}
\end{eqnarray}

Since the vectors $\mathbf{t}$ and $\mathbf{\tau }$ verify the Hyllerras
commutation relations [17] 
\begin{eqnarray}
\left[ t,\mathbf{t}\cdot \mathbf{\tau }\right] &=&it,  \TCItag{42.a} \\
\left[ t,\tau ^{2}\right] &=&\left( 2i\mathbf{\tau }\cdot \mathbf{t}%
-1\right) \dfrac{1}{t},  \TCItag{42.b}
\end{eqnarray}%
then, taking into account (42.b), the left-side of (41) is expanded following%
\begin{eqnarray}
t\left( 1+\tau ^{2}\right) t\left( 1+\tau ^{2}\right) \psi \left( t,\Phi
\right) &\equiv &\left( t^{2}+t\tau ^{2}t\right) \left( 1+\tau ^{2}\right)
\psi \left( t,\Phi \right)  \nonumber \\
&=&\left[ t^{2}+t^{2}\tau ^{2}-2it\mathbf{\tau }\cdot \mathbf{t}\frac{1}{t}+1%
\right] \left( 1+\tau ^{2}\right) \psi \left( t,\Phi \right) ,  \TCItag{43}
\end{eqnarray}%
and using (42.a), we obtain%
\begin{equation}
\left[ t^{2}\left( 1+\tau ^{2}\right) -2i\mathbf{\tau }\cdot \mathbf{t}+3%
\right] \left( 1+\tau ^{2}\right) \psi \left( t,\Phi \right) =\left(
2a+1\right) ^{2}\psi \left( t,\Phi \right) .  \tag{44}
\end{equation}

Proceeding now to canonical transformations, which permit to exchange the
coordinates to momenta through%
\begin{equation}
\left( 
\begin{array}{c}
\mathbf{t}^{\prime } \\ 
\mathbf{\tau }^{\prime }%
\end{array}%
\right) =%
\begin{pmatrix}
0 & 1 \\ 
-1 & 0%
\end{pmatrix}%
\left( 
\begin{array}{c}
\mathbf{t} \\ 
\mathbf{\tau }%
\end{array}%
\right) .  \tag{45}
\end{equation}

It is clear that the transformation (45) has the following properties $%
t^{\prime 2}=\tau ^{2},\ \tau ^{\prime 2}=t^{2}\ $and $\mathbf{\tau }\cdot 
\mathbf{t}=-\mathbf{t}^{\prime }\cdot \mathbf{\tau }^{\prime }$. These
transformations amount to describe a two-dimensional Fourier transform [1].
In terms of these, (44) becomes%
\begin{equation}
\left[ \tau ^{\prime 2}\left( 1+t^{\prime 2}\right) +2i\mathbf{t}^{\prime
}\cdot \mathbf{\tau }^{\prime }+3\right] \left( 1+\tau ^{\prime 2}\right)
\Psi \left( t^{\prime },\Phi ^{\prime }\right) =\left( 2a+1\right) ^{2}\Psi
\left( t^{\prime },\Phi ^{\prime }\right) ,  \tag{46}
\end{equation}%
where, as a consequence of this, the wave\ function $\Psi \left( t^{\prime
},\Phi ^{\prime }\right) $\ is the Fourier transform of $\psi \left( t,\Phi
\right) $.

Using coordinates $t^{\prime }$\ and $\Phi ^{\prime }$, the Fourier
transform of (40) reads%
\begin{equation}
\tau ^{\prime 2}\equiv -\frac{1}{t^{\prime }}\partial _{t^{\prime }}\left(
t^{\prime }\partial _{t^{\prime }}\right) -\frac{1}{t^{\prime 2}}\partial
_{\Phi ^{\prime }}^{2}+\frac{2}{t^{\prime }}\partial _{t^{\prime
}}Q_{a,\Gamma }^{\left( M\right) }\left( t^{\prime }\right) .  \tag{47}
\end{equation}%
where, as was established for the wave functions above, the function $\dfrac{%
1}{t^{\prime }}\partial _{t^{\prime }}Q_{a,\Gamma }^{\left( M\right) }\left(
t^{\prime }\right) $\ is the Fourier transform of $\dfrac{1}{t}\partial
_{t}q_{a,\Gamma }^{\left( M\right) }\left( t\right) $\ (see appendix).

Inserting both (47) and definition of scalar product $\mathbf{t}^{\prime
}\cdot \mathbf{\tau }^{\prime }=-it^{\prime }\partial _{t^{\prime }}$ into
(46), we obtain%
\begin{gather}
\left[ \left( -\frac{1}{t^{\prime }}\partial _{t^{\prime }}\left( t^{\prime
}\partial _{t^{\prime }}\right) -\frac{1}{t^{\prime 2}}\partial _{\Phi
^{\prime }}^{2}+\frac{2}{t^{\prime }}\partial _{t^{\prime }}Q_{a,\Gamma
}^{\left( M\right) }\left( t^{\prime }\right) \right) \left( 1+t^{\prime
2}\right) +2t^{\prime }\partial _{t^{\prime }}+3\right] \left( 1+t^{\prime
2}\right) \Psi \left( t^{\prime },\Phi ^{\prime }\right)  \nonumber \\
=\left( 2a+1\right) ^{2}\Psi \left( t^{\prime },\Phi ^{\prime }\right) . 
\tag{48}
\end{gather}

By introducing a new wave function $\xi \left( t^{\prime },\Phi ^{\prime
}\right) \footnote{%
It is evident that the function $\xi \left( t^{\prime },\Phi ^{\prime
}\right) $ is the same wave function introduced in subsection 2.2.}$\
according to [1]%
\begin{equation}
\xi \left( t^{\prime },\Phi ^{\prime }\right) =\left( 1+t^{\prime 2}\right)
^{3/2}\Psi \left( t^{\prime },\Phi ^{\prime }\right) ,  \tag{49}
\end{equation}%
we obtained from (48), after derivation and a long calculation%
\begin{equation}
\left[ -\frac{1}{t^{\prime }}\partial _{t^{\prime }}\left( t^{\prime
}\partial _{t^{\prime }}\right) -\frac{1}{t^{\prime 2}}\partial _{\Phi
^{\prime }}^{2}+\frac{2}{t^{\prime }}\partial _{t^{\prime }}Q_{a,\Gamma
}^{\left( M\right) }\left( t^{\prime }\right) \right] \xi \left( t^{\prime
},\Phi ^{\prime }\right) =\frac{4a\left( a+1\right) }{\left( 1+t^{\prime
2}\right) ^{2}}\xi \left( t^{\prime },\Phi ^{\prime }\right) .  \tag{50}
\end{equation}

We introduce the wave function $\xi \left( t^{\prime },\Phi ^{\prime
}\right) =\mathcal{U}\left( t^{\prime }\right) \func{e}^{im\Phi ^{\prime }}$
in (50). At this point, it is interesting to re-express the second term in
the left-hand side of (50) in term of the second-order differential
operator, as was done in (35), accordingly to%
\begin{equation}
\partial _{\Phi ^{\prime }}^{2}\equiv E_{M}-a^{2},  \tag{51}
\end{equation}%
which finally brings (50) to%
\begin{equation}
\left[ -\frac{1}{t^{\prime }}\partial _{t^{\prime }}\left( t^{\prime
}\partial _{t^{\prime }}\right) -\frac{E_{M}-a^{2}}{t^{\prime 2}}+\frac{2}{%
t^{\prime }}\partial _{t^{\prime }}Q_{a,\Gamma }^{\left( M\right) }\left(
t^{\prime }\right) \right] \mathcal{U}\left( t^{\prime }\right) =\frac{%
4a\left( a+1\right) }{\left( 1+t^{\prime 2}\right) ^{2}}\mathcal{U}\left(
t^{\prime }\right) ,  \tag{52}
\end{equation}%
Since the Schr\"{o}dinger equation deduced in (52) is the same as previously
derived in (33), then 
\begin{eqnarray}
\tciFourier \left[ \frac{1}{t}\partial _{t}q_{a,\Gamma }^{\left( M\right)
}\left( t\right) \right] &=&\frac{1}{t^{\prime }}\partial _{t^{\prime
}}Q_{a,\Gamma }^{\left( M\right) }\left( t^{\prime }\right)  \nonumber \\
&\equiv &\frac{1}{t^{\prime }}\partial _{t^{\prime }}q_{\mu ,\Gamma
}^{\left( PT\right) }\left( t^{\prime }\right) ,  \TCItag{53}
\end{eqnarray}%
where $\tciFourier $ is the Fourier transform operator. It should be clear
that both generalized potentials are connected by means of the Hankel
transform which is one in a large number of ways in which the Fourier
transform can be written (see Appendix, (A.14)).

Identifying (33) to (52), term by term, is considered as an alternative way
able to reproduce the relationship connecting both energy spectra. Then,
comparing both second terms and right-hand sides of (33) and (52),
respectively, and taking into consideration $a=\lambda -\frac{1}{2}$, we
obtain%
\begin{equation}
E_{PT}=E_{M}+\lambda -\mu -\frac{1}{2}.  \tag{54}
\end{equation}

We note that the energy spectrum of the generalized P\"{o}schl-Teller
potential is related to the energy spectrum of the generalized Morse
potential by shifting the $E_{M}$\ value by $\lambda -\mu -\frac{1}{2}$.

\section{Conclusion}

The main purpose of the present paper is to extend the procedure of [1] for
generalized Morse and P\"{o}schl-Teller potentials in order to connect them
with the Schr\"{o}dinger equation. These generalized potentials are
determined by applying the generalized version of creation and annihilation
operators provided by the factorization method. Our primary concern is to
construct mathematically the Schr\"{o}dinger equation for generalized P\"{o}%
schl-Teller potential starting off with the Schr\"{o}dinger equation for
Morse potential. It is found that both Schr\"{o}dinger equations are related
by the Fourier transform. Our secondary purpose consists to offer an
explanation that there exist an intimate correlation between generalized
potentials with a view to establishing a causal connection underlying the
relationship which connects them. This establishment of the correspondence
between the generalized potentials means that the known generalized Morse
potential automatically provides us with the generalized P\"{o}schl-Teller
potential and vice versa, through the Hankel transform (see (A.14)). As a
consequence of this, energy spectrum of generalized P\"{o}schl-Teller
potential is related to the energy spectrum of generalized Morse potential
by shifting the $E_{M}$\ value by $\lambda -\mu -\dfrac{1}{2}$.

It will be interesting to analyze the connection between generalized Morse
and P\"{o}schl-Teller potentials using path-integrals formalism rather than
Schr\"{o}dinger equation, where Refs. [18,19] suggests a close connection
between the eigenfunctions and sum-over-paths representation of the
propagators.

\section{Appendix}

In this appendix, we add some mathematical details to the discussion of
section 3 in order to prove that both generalized potentials are linked by
Hankel transform. Comparing (33) and (50), and taking into account (49), we
can write that 
\begin{subequations}
\label{A.1}
\begin{eqnarray}
\partial _{t^{\prime }}Q_{a,\ \Gamma }^{\left( M\right) }\left( t^{\prime
}\right) \xi \left( t^{\prime },\Phi ^{\prime }\right) &=&\left( 1+t^{\prime
2}\right) ^{3/2}\partial _{t^{\prime }}Q_{a,\Gamma }^{\left( M\right)
}\left( t^{\prime }\right) \Psi \left( t^{\prime },\Phi ^{\prime }\right) 
\nonumber \\
&=&\left( 1+t^{\prime 2}\right) ^{3/2}\tciFourier \left[ \partial _{t}q_{\mu
,\Gamma }^{\left( PT\right) }\left( t\right) \psi \left( t,\Phi \right) %
\right] .  \TCItag{A.1}
\end{eqnarray}

The Fourier transform for the function of two variables, e.g. $t_{x}$ and $%
t_{y}$\ of section 3 is 
\end{subequations}
\begin{equation}
\Psi \left( t_{x},t_{y}\right) =\dint\limits_{-\infty }^{\infty
}\dint\limits_{-\infty }^{\infty }dt_{x}dt_{y}\psi \left( t_{x},t_{y}\right)
\exp \left[ -i\left( t_{x}t_{u}+t_{y}t_{v}\right) \right] .  \tag{A.2}
\end{equation}

Let us make a change of variables taken above in (37), (A.2) becomes%
\begin{eqnarray}
\Psi \left( t^{\prime },\Phi ^{\prime }\right) &=&\dint\limits_{0}^{2\pi
}d\Phi \dint\limits_{0}^{\infty }tdt\psi \left( t,\Phi \right) \exp \left[
-i\left( tt^{\prime }\cos \Phi \cos \Phi ^{\prime }+tt^{\prime }\sin \Phi
\sin \Phi ^{\prime }\right) \right]  \nonumber \\
&=&\dint\limits_{0}^{2\pi }d\Phi \dint\limits_{0}^{\infty }tdt\psi \left(
t,\Phi \right) \func{e}^{-\ itt^{\prime }\cos \left( \Phi -\Phi ^{\prime
}\right) }  \TCItag{A.3}
\end{eqnarray}

Since $\psi \left( t,\Phi \right) =\mathcal{R}\left( t\right) \func{e}%
^{im\Phi }$ and $\xi \left( t^{\prime },\Phi ^{\prime }\right) =\mathcal{U}%
\left( t^{\prime }\right) \func{e}^{im\Phi ^{\prime }}$, (A.1) becomes%
\begin{equation}
\partial _{t^{\prime }}Q_{a,\Gamma }^{\left( M\right) }\left( t^{\prime
}\right) \xi \left( t^{\prime },\Phi ^{\prime }\right) =\left( 1+t^{\prime
2}\right) ^{3/2}\dint\limits_{0}^{2\pi }d\Phi \dint\limits_{0}^{\infty }tdt%
\mathcal{R}\left( t\right) \partial _{t}q_{\mu ,\Gamma }^{\left( PT\right)
}\left( t\right) \func{e}^{-\ itt^{\prime }\cos \left( \Phi -\Phi ^{\prime
}\right) +im\Phi }.  \tag{A.4}
\end{equation}

We perform now the $\Phi $-integration. To this end, the useful way of
treating this integration is to employ the integral representation of the
cylindrical Bessel functions [20,21]%
\begin{equation}
J_{\upsilon }\left( x\right) \equiv \frac{1}{2\pi i}\doint dz\
z^{-1-\upsilon }\exp \left[ \frac{x}{2}\left( z-\frac{1}{z}\right) \right] .
\tag{A.5}
\end{equation}

Let $z=e^{i\Phi }$\ where the path of integration is the unit circle, the
integral (A.5) becomes, by setting $\upsilon =m$\ and $x=tt^{\prime }$%
\begin{equation}
J_{m}\left( tt^{\prime }\right) =\frac{1}{2\pi }\dint\limits_{0}^{2\pi
}d\Phi \func{e}^{itt^{\prime }\sin \Phi -im\Phi }.  \tag{A.6}
\end{equation}

In order to recover the exponent in (A.4), a new change of variable, $\Phi
\rightarrow \Phi -\Phi ^{\prime }-\pi /2,$\ is introduced%
\begin{equation}
J_{m}\left( tt^{\prime }\right) =\frac{1}{2\pi }i^{m}\func{e}^{im\Phi
^{\prime }}\dint\limits_{0}^{2\pi }d\Phi \func{e}^{-\ itt^{\prime }\cos
\left( \Phi -\Phi ^{\prime }\right) -im\Phi },  \tag{A.7}
\end{equation}%
and by replacing $m\rightarrow -m$\ and taking into account the identity $%
J_{-m}\left( x\right) =\left( -1\right) ^{m}J_{m}\left( x\right) $\ [22,23],
we find that%
\begin{equation}
\dint\limits_{0}^{2\pi }d\Phi \func{e}^{-\ itt^{\prime }\cos \left( \Phi
-\Phi ^{\prime }\right) +im\Phi }=2\pi \left( -i\right) ^{m}\func{e}^{im\Phi
^{\prime }}J_{m}\left( tt^{\prime }\right) .  \tag{A.8}
\end{equation}

Inserting (A.8) into (A.4), we obtain%
\begin{equation}
\partial _{t^{\prime }}Q_{a,\Gamma }^{\left( M\right) }\left( t^{\prime
}\right) \mathcal{U}\left( t^{\prime }\right) =2\pi \left( -i\right)
^{m}\left( 1+t^{\prime 2}\right) ^{3/2}\dint\limits_{0}^{\infty }tdt\mathcal{%
R}\left( t\right) \partial _{t}q_{\mu ,\Gamma }^{\left( PT\right) }\left(
t\right) J_{m}\left( tt^{\prime }\right) .  \tag{A.9}
\end{equation}

It was already established by Alhassid et al. [1] that the connection
between the $"$radial$"$ solutions of the Schr\"{o}dinger equation with
Morse and P\"{o}schl-Teller potentials is%
\begin{equation}
\mathcal{U}\left( t^{\prime }\right) =2\pi \left( -i\right) ^{m}\left(
1+t^{\prime 2}\right) ^{3/2}\dint\limits_{0}^{\infty }tdt\mathcal{R}\left(
t\right) J_{m}\left( tt^{\prime }\right) .  \tag{A.10}
\end{equation}

Substituting (A.10) into (A.9), we obtain after simplification%
\begin{equation}
\partial _{t^{\prime }}Q_{a,\Gamma }^{\left( M\right) }\left( t^{\prime
}\right) \dint\limits_{0}^{\infty }tdt\mathcal{R}\left( t\right) J_{m}\left(
tt^{\prime }\right) =\dint\limits_{0}^{\infty }tdt\mathcal{R}\left( t\right)
\partial _{t}q_{\mu ,\Gamma }^{\left( PT\right) }\left( t\right) J_{m}\left(
tt^{\prime }\right) .  \tag{A.11}
\end{equation}

Performing now the $t^{\prime }$-integration and by interchanging the order
of integration, we find that%
\begin{align}
\dint\limits_{0}^{\infty }tdt\mathcal{R}\left( t\right)
\dint\limits_{0}^{\infty }dt^{\prime }t^{\prime }\frac{1}{t^{\prime }}%
\partial _{t^{\prime }}Q_{a,\Gamma }^{\left( M\right) }\left( t^{\prime
}\right) J_{m}\left( tt^{\prime }\right) & =\dint\limits_{0}^{\infty }tdt%
\mathcal{R}\left( t\right) \partial _{t}q_{\mu ,\Gamma }^{\left( PT\right)
}\left( t\right) \dint\limits_{0}^{\infty }dt^{\prime }J\left( tt^{\prime
}\right)  \nonumber \\
& =\dint\limits_{0}^{\infty }tdt\mathcal{R}\left( t\right) \partial
_{t}q_{\mu ,\Gamma }^{\left( PT\right) }\left( t\right) \frac{1}{t}, 
\tag{A.12}
\end{align}%
where the integral of Bessel function depending on a parameter $p$\ yields
[20,21]%
\begin{equation}
\dint\limits_{0}^{\infty }dxJ_{\upsilon }\left( px\right) =\frac{1}{p}. 
\tag{A.13}
\end{equation}

Therefore, by identifying terms in (A.12) and using the Hankel transform
(Fourier-Bessel integral) [20,21] which is one in a large number of ways in
which the Fourier transform can be written, we obtain

\begin{eqnarray}
\frac{1}{t}\partial _{t}q_{\mu ,\Gamma }^{\left( PT\right) }\left( t\right)
&=&\dint\limits_{0}^{\infty }t^{\prime }dt^{\prime }\frac{1}{t^{\prime }}%
\partial _{t^{\prime }}Q_{a,\Gamma }^{\left( M\right) }\left( t^{\prime
}\right) J_{m}\left( tt^{\prime }\right)  \nonumber \\
&=&\tciFourier \left[ \frac{1}{t^{\prime }}\partial _{t^{\prime
}}Q_{a,\Gamma }^{\left( M\right) }\left( t^{\prime }\right) \right] , 
\TCItag{A.14}
\end{eqnarray}%
which is a result already established above in (53).

\end{document}